**Title:**

# Critical periods and Autism Spectrum Disorders, a role for sleep

**Authors:**

Elizabeth Medina[1], Sarah Peterson[1], Kristan Singletary[1], Lucia Peixoto[1]*.

**Affiliations:**

1. Department of Translational Medicine and Physiology. Sleep and Performance Research Center. Elson S. Floyd College of Medicine, Washington State University, Spokane, Washington, United States

* To whom correspondence should be addressed: Pharmaceutical and Biomedical Science Building #220, 412 E. Spokane Falls Blvd. Spokane WA 99202. Tel: 509-368-6764, Fax: 509-368-7882. Email: lucia.peixoto@wsu.edu.

**Keywords**: Autism, Sleep, synaptic plasticity, critical periods of development.

**SUPPORT**

This work was supported by the K01NS104172 from NIH/NINDS to Peixoto L.

## Abstract

Brain development relies on both experience and genetically defined programs. Time windows where certain brain circuits are particularly receptive to external stimuli, resulting in heightened plasticity, are referred to as "critical periods". Sleep is thought to be essential for normal brain development. Importantly, studies have shown that sleep enhances critical period plasticity and promotes experience-dependent synaptic pruning in the developing mammalian brain. Therefore, normal plasticity during critical periods depends on proper sleep. Problems falling and staying asleep occur at a higher rate in Autism Spectrum Disorder (ASD) relative to typical development. In this review, we explore the potential link between sleep, critical period plasticity, and ASD. First, we review the importance of critical period plasticity in typical development and the role of sleep in this process. Next, we summarize the evidence linking ASD with deficits in synaptic plasticity in rodent models of high-confident ASD gene candidates. We then show that almost all the high-confidence rodent models of ASD that show sleep deficits also display plasticity deficits. Given how important sleep is for critical period plasticity, it is essential to understand the connections between synaptic plasticity, sleep, and brain development in ASD. However, studies investigating sleep or plasticity during critical periods in ASD mouse models are lacking. Therefore, we highlight an urgent need to consider developmental trajectory in studies of sleep and plasticity in neurodevelopmental disorders.

## Significance statement

In this review, we support the known link between the genetic basis of ASD and deficits in synaptic plasticity and highlight that there is a considerable overlap between deficits in sleep and deficits in plasticity in ASD mouse models. Given an established role of sleep and plasticity during critical periods, investigating both early in life may be key to understanding long-term consequences of abnormal sleep on synaptic plasticity in both typical and atypical brain development.

## Introduction

Brain development relies on both experience and genetically defined programs. Brain circuits are established through neuronal proliferation and migration. Experience remodels circuits through both synaptogenesis and pruning, a process referred to as synaptic plasticity[1]. Time windows where certain brain circuits are particularly receptive to external stimuli, resulting in heightened plasticity, are referred to as "critical periods"[2–4]. While heightened plasticity allows us to shape neuronal circuits, this also introduces vulnerability. Therefore, disruptions early in life can result in neuronal circuits that respond differently to experiences later on[3–6]. In the last four decades, developmental windows for our brain's ability to attune to experience have been widely studied, emphasizing the importance of critical periods in the developing brain[3]. During early postnatal development, and while critical periods occur, young mammals spend most of their time sleeping. Sleep is thought to be essential for normal brain development[7–11]. Importantly, studies have shown that sleep enhances critical period plasticity[12] and promotes experience-dependent synaptic pruning in the developing cortex[13,14]. Therefore, normal plasticity during critical periods depends on proper sleep.

Problems falling and staying asleep occur at a higher rate in individuals with Autism Spectrum Disorder (ASD) than in individuals with typical development, affecting up to 86% of individuals[15]. Sleep problems in individuals with ASD are predictive of the severity of ASD core symptoms[16] and lead to a reduced quality of life of individuals and caregivers[17]. Although sleep problems are often thought of as a co-occurring condition, the fundamental role of sleep in brain development points to sleep problems as a potential core factor in ASD. Recent studies have shown that problems falling asleep in the first year of life precede the ASD diagnosis and are associated with hippocampal overgrowth in infants at high-risk for ASD[18]. Recent studies have also shown that problems falling asleep predict impairments in behavior regulation later in children with ASD[19]. However, studies investigating the mechanisms that link sleep early in life and ASD are lacking. ASD is diagnosed behaviorally on average between 4-5 years of age based on symptoms in two core domains, social deficits and repetitive behavior[20]. Altered trajectories of development in individuals that go on to be diagnosed with ASD can be detected earlier in life[21]. For example, over-expansion of the cortical surface area between 6 and 12 months of age and subsequent brain volume overgrowth is predictive of later ASD diagnosis in high-risk infants[22]. The high rate of cortical surface and brain volume expansion that occurs after birth in mammals is thought to facilitate the contributions of postnatal experience[23]. Therefore, alterations in cortical and brain growth early in life can affect how experience molds the brain. The rapid brain growth in the postnatal years is accompanied by high rates of synaptogenesis and pruning, fundamental substrates of synaptic plasticity. Deficits in plasticity early in life are thought to be important contributors to the ASD phenotype[16,17,24]. Given the relationship between sleep and plasticity, it is reasonable to propose that sleep and plasticity deficits during critical periods of development are inextricably linked in ASD.

In this review, we explore the potential link between critical period plasticity, sleep, and ASD. First, we review the importance of critical period plasticity in typical development and the role of sleep in this process. Next, we summarize the evidence linking ASD with deficits in synaptic plasticity in mouse models of high-confident ASD gene candidates, and we show that almost all the high-confidence mouse models of ASD that show sleep deficits also have been shown to display plasticity deficits. Given how important sleep is in early life, and for critical period plasticity in particular, this review highlights the importance of understanding connections between synaptic plasticity, sleep, and brain development in ASD.

## The importance of critical periods in typical brain development

The brain's typical development relies on multiple developmental processes, such as cell proliferation, migration, circuit formation and maturation, all driven by genetic and environmental stimuli[2]. Experience promotes rewiring and remodeling through synaptogenesis and pruning, often referred to as a process called synaptic plasticity[1]. Synaptic plasticity refers to the ability of synapses to strengthen or weaken in response to use or disuse. Critical periods of development, in which brain circuits are particularly plastic, have been recognized for nearly a century[25]. It is important to note that the critical period of plasticity is simply an extreme form of plasticity; while this is elevated in early development, different types of plasticity can occur throughout our lifespan. While the timeline for brain development differs across species, critical periods have been established in both humans and animals. Critical periods also differ in timing based on input modality. During postnatal development, critical periods occur at ages in which other important processes for brain development also occur at a high rate, in particular synaptogenesis and pruning.

The visual system has long been favored as a model for critical period plasticity because experience strongly regulates the visual cortex's maturation[25,26]. The best-studied example of critical period plasticity is the ocular dominance plasticity model (ODP) in the visual cortex as this model allows for measures of neuronal activity *in vivo* in unanesthetized brains in a system whose cellular processes are well established. In 1959, Hubel and Wiesel discovered that specific neurons in a cat's primary visual cortex were activated through stimuli presented to one eye or the other, termed ocular dominance[27,28]. They used monocular deprivation (MD) to deprive one eye of visual experience, which caused a decrease in the cortical neurons' response associated with the deprived eye and increased response to neurons associated with the spared eye in the binocular cortex. MD in a young kitten resulted in the irreversible loss of responsiveness in those cells even though there was no damage to the retina[29–31]. In contrast, the same MD in an adult cat did not result in a loss of visual acuity. Importantly, this critical period is not a simple age-dependent mechanism, as animals that are only exposed to complete darkness in early life have a delayed onset of this critical period[32,33]. Therefore, the critical period for ODP is both age and experience dependent.

The consolidation of ODP is divided into three stages. In the first stage, there is a decrease in response to the deprived eye, which is hypothesized to result from the loss of deprived-eye connections or depression in their synaptic efficacy. The dramatic loss of response from the deprived eye is thought to occur through Hebbian plasticity. Hebbian plasticity is defined as changes in the connection strength (either strengthening or weakening) between two neurons resulting from correlated firing[34]. Long-term potentiation (LTP) and long-term depression (LTD) are two forms of Hebbian plasticity. In the second stage of ODP, there is an increase of response in both eyes, but most notably the open eye. This stage is thought to operate through both homeostatic and Hebbian plasticity. Homeostatic plasticity, such as synaptic scaling, is a form of plasticity that attempts to stabilize neuron or neuronal circuits' activity in the event of a perturbation, such as a change in synapse number or strength that may alter excitability. The ODP model has been used to demonstrate the role of sleep on critical period plasticity as described below.

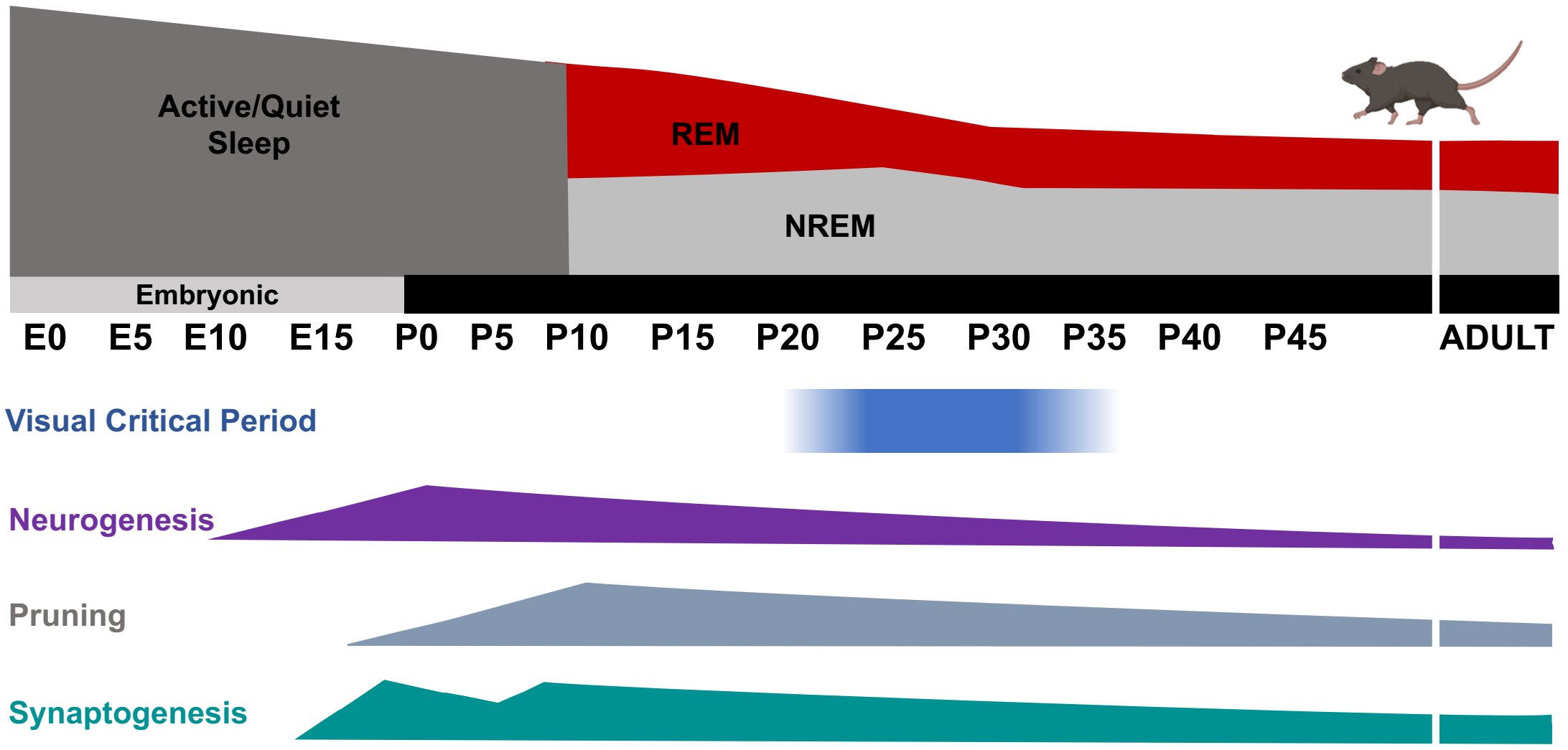

**Figure 1. Timeline of sleep activity and developmental time points in the mouse brain.** Rectangles are estimates of start and end time of visual critical period[25]. Triangles indicate the onset and peak of developmental periods[26]. Abbreviation REM= rapid eye movement, NREM= nonrapid eye movement E= embryonic, P= postnatal.

## The role of sleep in the consolidation of plasticity

During early postnatal development, and while critical periods occur, mammals spend most of their time sleeping. The amount of sleep changes as an organism develops. Sleep states in mammals are traditionally defined using electroencephalographic recordings (EEG) of rapid-eye-movement (REM) sleep and non-rapid-eye movement (NREM) sleep. REM and NREM do not emerge in rodents until postnatal (P) days 12-16 and are preceded by relatively undifferentiated states termed active sleep and quiet sleep which are mostly behaviorally defined[35]. **Figure 1** shows how critical events in postnatal brain development relate in timing to the visual critical period and the development of sleep states defined by brain electrical activity in mice. Overall, these events are highly synchronous during early life.
While the function of sleep is not fully understood, several functional and morphological studies *in vivo* provide direct evidence that sleep is important for synaptic plasticity during development and into adulthood[36]. The mechanisms underlying this relationship are less clear.

Initial studies have shown that sleep enhances plasticity during the critical period for ODP in the visual cortex of kittens[12]. In addition, REM sleep has been shown to be required for phosphorylation of several kinases involved in synaptic plasticity following MD, such as ERK[37–39] . Therefore, REM sleep enhances plasticity by influencing phosphorylation events in established signaling pathways. In young mice, REM sleep has been shown to be required for the reduction of input from the deprived eye following MD. It has also been shown to be required for synaptic pruning which is thought to underlie this process[13,14]. **Figure 2** depicts the ODP model and how sleep or sleep deprivation can influence experience-dependent plasticity following. In the visual cortex under normal conditions neuronal activity represents predominantly stimuli from the contralateral eye (**2A**). Following MD, activity shifts towards input from the non-deprived eye (**2B**, experience-dependent plasticity). That shift occurs when the animal is allowed to sleep following MD (**2C**), but not if the animal is sleep deprived (**2D**). Therefore, sleep is necessary for plasticity in the ODP model. This is supported by multiple lines of evidence, which show that sleep can both strengthen and weaken different synapses in response to experience during wake[12,13,36,37,40]. Sleep has also been shown to have an important role in memory consolidation, which requires synaptic plasticity[41]. Overall, these studies provide evidence that sleep, and in particular REM sleep, are essential for the consolidation of plasticity, especially during critical periods of development. Therefore, sleep alterations during early life may have long lasting effects on experience-dependent synapse remodeling.

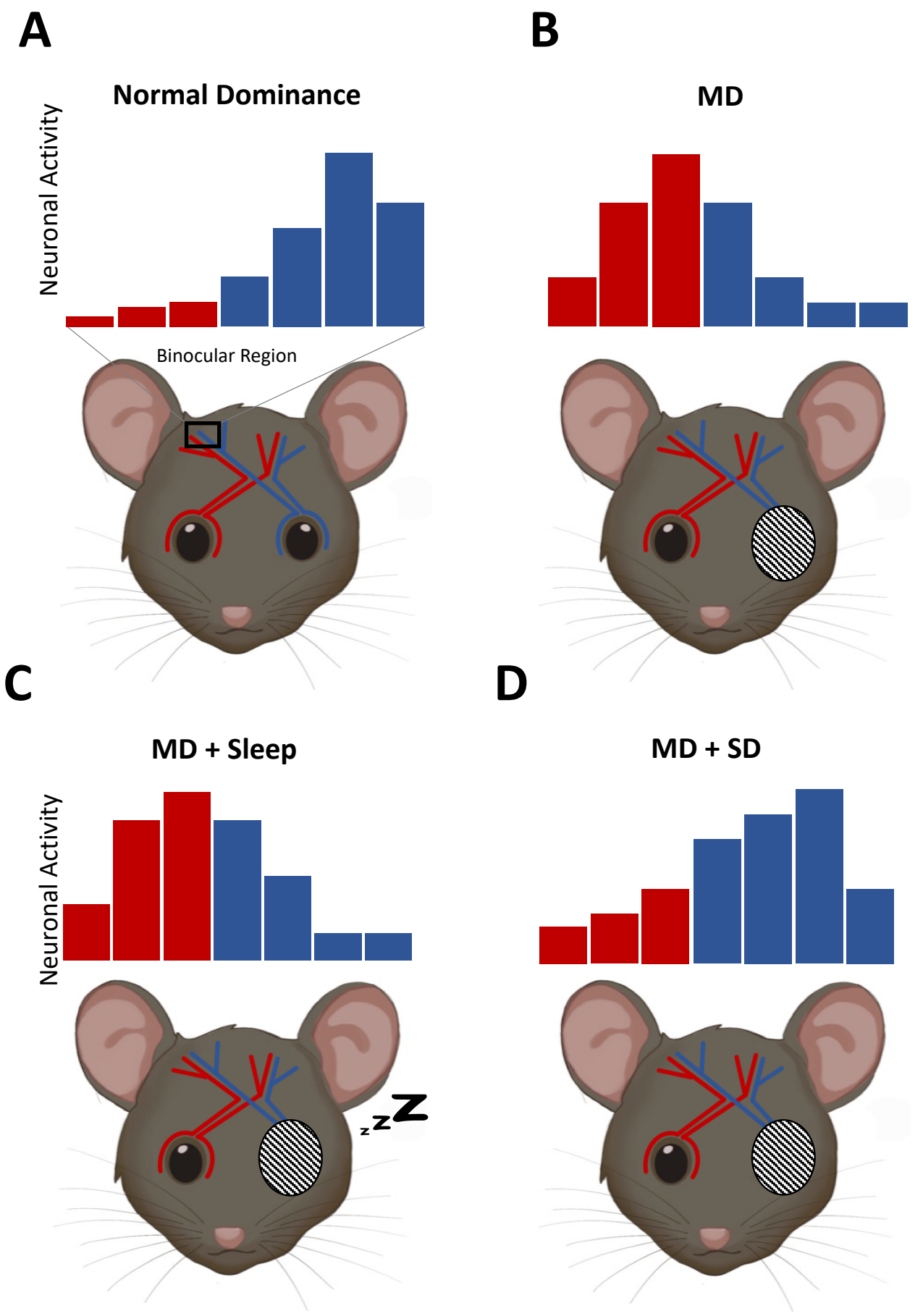

**Figure 2. The Ocular Dominance Plasticity model. A)** In the binocular region of the visual cortex most neurons respond to input from the contralateral eye while very few respond to input from the ipsilateral eye input only. **B)** Monocular deprivation (MD) during this critical period results in more neurons responding to the non-deprived eye and less input from the contralateral, deprived eye**. C)** Sleep is necessary to consolidate the effects of MD**. D)** When sleep deprivation (SD) follows MD, the shift in response is not seen.

## Sleep problems in Autism Spectrum Disorders

In ASD, sleep problems occur at a much higher rate than in typical development, affecting up to 86% of individuals with ASD[42–44]. ASD is the most prevalent neurodevelopmental disorder in the US. Three key features characterize ASD sleep problems: problems falling asleep, problems staying asleep, and overall less sleep[44]. Sleep problems predict the severity of ASD core diagnostic symptoms and associated challenging behaviors[42], and often increase with age[44], which imposes a severe burden on caregivers. Recent studies using parent reports of sleep on a cohort of baby siblings of individuals with ASD (who are at a higher risk for a diagnosis) found that problems falling asleep in the first year of life precede an autism diagnosis and are associated with altered patterns of brain development, specifically hippocampal overgrowth[45]. Baby-sibling studies have also shown that sleep problems in autistic children as young as four years old are associated with increased 'higher-order' restricted and repetitive behaviors later in childhood[46]. This suggests that sleep problems early in life are predictive of symptom severity across the lifespan. This view is supported by recent studies showing that problems falling asleep predict impairments in behavior regulation later in children with ASD[19]. However, whether there is a causal mechanism linking sleep problems and brain dysfunction during development remains unknown.

There are several monogenic syndromes associated with ASD. The recapitulation of these syndromes in animal models has provided insights into mechanisms associated with the disorder. Genes linked to ASD and available mouse models can be found in the Simons Foundation Autism Research Initiative (SFARI) gene database. In that database genes are categorized according to the strength of evidence linking them to ASD based on human genetic studies as follows: S (Syndromic), 1 (High Confidence), 2 (Strong Candidate), 3 (Suggestive Evidence), 4 (Minimal Evidence), 5 (Hypothesized). Few of the mouse ASD models that have been studied in regards to sleep recapitulate any features of clinical sleep phenotype, which are: reduced sleep time, sleep fragmentation, and increased latency to fall asleep[47]. Only 15 ASD mouse models display reduced sleep, of which the following 10 are high-confidence or syndromic: *Shank3, Fmr1, Mecp2, Ube3a, Rims1, Scn1a, Camk2a, Cacna1c, Scn8a* and CNV 16p11.2. While the *Shank3, Ube3a, and Scn1a* mouse models all display sleep fragmentation, the *Shank3*$^{\Delta C}$ mouse model is the only high-confidence ASD mouse model to also display increased sleep latency.

Sleep patterns across development have been investigated in only one mouse model, *Shank3*$^{\Delta C}$ mice. Studies show that *Shank3*$^{\Delta C}$ mice sleep less throughout their lifespan and show reduced NREM sleep but paradoxically display dramatically increased levels of REM sleep at postnatal day 23[48]. At this age mice have achieved 70% of their maximal brain growth and are roughly equivalent to 9-10 month-old human infants based on brain volume[49]. *Shank3*$^{\Delta C}$ mice at P23 also show EEG spectral signatures during wake that may be indicative of cortical overgrowth[48]. This time in rodents coincides with critical periods when the brain is exquisitely sensitive to changes in the environment (**Figure 1**). Therefore, studies in young mice, although limited, support the hypothesis from infant studies that sleep may be altered early in life and be a core feature of the ASD phenotype. In turn this predicts that sleep problems early in life may be highly correlated and perhaps predictive of synaptic plasticity deficits, which are well documented in ASD[50].

**Deficits in synaptic plasticity in ASD**

ASD encompasses a set of highly heterogenous neurodevelopmental disorders that share common behavioral abnormalities. Despite the genetic complexity of ASD, genes associated with ASD are often related to synaptic function and transcription regulation. To establish the number of ASD mouse models that have also been associated with deficits in synaptic plasticity we reviewed studies within the SFARI gene database (January 2021 release). The original dataset included copy number variants (CNV) and genes that were either syndromic (S) or had an evidence score of 1 (high confidence) and encompassed 105 genes and 438 related articles.  Based on primary research articles we determined which genes or genomic regions influenced plasticity. We defined an effect on plasticity based on whether a mutation in the gene or a CNV produced one or more of the following phenotypes in mice: 1) change in dendritic spine morphology or number, 2) change in plasticity measured by electrophysiology, or 3) change in learning/memory behavioral task. The results are displayed in **Table 1** and include 4 CNVs and 65 genes.

While ASD encompasses a wide range of neurodevelopmental disorders, identifying common pathways may aid the understanding and treatment of ASD core features. ASD risk genes identified so far belong to convergent biological pathways relating to transcription, chromatin structure, and synaptic function. Although the earliest ASD risk genes to be identified were known to function at the synapse, it has become clear that a distinct subset of ASD genes are chromatin-modifiers and RNA-binding proteins. Of the 65 ASD risk genes we identified to have a role in synaptic plasticity based on animal model studies, 14 are associated with the cell membrane and known to function at the synapse *(e.g.: Ctnnb1, Dscam, Shank2, Shank3, Ank2, Camk2a, Dlg4, Dmd, Fmr1, Gria2, Grin2b, Nrxn1, Nlgn3, Rims1).* In addition, a large fraction of genes in Table 1 act in the nucleus as transcription/chromatin regulators *e.g., Pogz, Ube3a, Nr3c2, Foxp2, Adnp, Mecp2, Chd2, Foxp1, Tcf4, Fmr1, Tsc1*. Some genes have both a role at the synapse and in the nucleus (*Ctnnb1, Shank3*). This likely reflects the fact that synaptic plasticity requires changes at the synapse to be stabilized through changes in transcription, translation, and epigenetic modifications[51].

**Abnormal sleep in ASD and its potential impact on plasticity**

Abnormal sleep in ASD and deficits in synaptic plasticity are not mutually exclusive phenotypes. Currently there are 10 high-confidence ASD mouse models that have been investigated using objective measures of sleep and display reduction in sleep time[47]. Mutations in 9 of those genes (16p11.2, *Fmr1, Shank3, Mecp2, Ube3a, Cacna1c, Rims1, Scn1a, Camk2a*) also lead to alterations in synaptic plasticity (**Table 1**). Therefore, most high-confidence ASD genes that affect sleep also influence synaptic plasticity. These genes that are both synaptic and nuclear is not surprising given the known role of sleep in the molecular pathways underlying long-term synaptic plasticity and memory formation[41]. Sleep has a profound effect on plasticity during neurodevelopment. Therefore, it is possible that in ASD a sleep phenotype may cause or contribute to deficits in synaptic plasticity from very early on. Nonetheless, very few studies have evaluated the role of ASD-associated genes on either sleep or plasticity early in life. The Shank3 ASD mouse is the only one in which sleep and plasticity have been examined and shown to be altered during critical periods of development[48,52]. There are currently no studies that have concurrently examined sleep and synaptic plasticity deficits in early life using an ASD mouse model. Given how important the early postnatal period has been shown to be for ASD, further examination of the interaction between sleep and plasticity during early life is necessary.

**Concluding remarks**

Animal models have been proven useful to understand mechanism underlying deficits in ASD. In this review, we support the known link between the genetic basis of ASD and deficits in synaptic plasticity and provide a comprehensive table of ASD mouse models that show plasticity deficits (**Table 1**). We also highlight that there is a considerable overlap between deficits in sleep, which are frequent in ASD, and deficits in plasticity in ASD mouse models. These suggest that the genetic basis is overlapping. This is not surprising given that sleep is important for the consolidation of plasticity. This role is particularly prominent during critical periods of brain development in which both plasticity and sleep are maximal. Recent studies suggest that sleep problems in ASD arise early in the postnatal period and are predictive of the severity of ASD core symptoms. Therefore, sleep problems: 1) may serve as an early biomarker for ASD and 2) may have functional consequences on plasticity during brain development. Further research is needed to understand the interaction between sleep and plasticity in ASD. Specifically, it is important to look early in life. Investigating sleep and plasticity during critical periods of development may be key to understanding long-term consequences of abnormal sleep on synaptic plasticity during both typical and atypical brain development.

**CONFLICT OF INTEREST STATEMENT**

Financial disclosure: The authors have no financial arrangements or connections to declare.
Non-financial disclosure: The authors have no conflicts of interests to declare.

**AUTHOR CONTRIBUTIONS**

All authors had full access to all the data in the study and take responsibility for the integrity of the data. *Conceptualization*, E.M. and L.P.; *Methodology*, E.M. and S.P.; *Data Curation,* E.M, S.P, and L.P; *Writing - Original Draft*, E.M.; *Writing - Review & Editing*, E.M, S.P., K.S., and L.P.; *Visualization*, E.M. and L.P.; *Supervision*, L.P.; *Funding Acquisition*, L.P.

**ACKNOWLEDGEMENTS**

We thank Dr. Marcos Frank for valuable discussions. This work was supported by K01NS104172 from NIH/NINDS to Peixoto L.

**DATA ACCESSIBILITY STATEMENT**

All data used in this review is available through the SFARI Gene Database.

## Tables

Table 1. High-Confidence ASD Genes and CNVs associated with alterations in neuroplasticity Summary of genes and CNVs extracted from the SFARI gene database (January 13th, 2021, release) for which mutations are associated with deficits in plasticity based on mouse model studies. Only copy number variants (CNV) or ASD genes that were either syndromic (S) or had an evidence score of 1 (high confidence) were considered. Column 1: gene symbol; column 2: brief description of the gene; column 3: PubMed ID associated articles supporting a role for the gene/CNV in neuroplasticity. “Yes” indicates a change in neuroplasticity reported in the literature in either: 1) dendritic spine morphology, 2) electrophysiological measures of plasticity (LTD, LTP, EPSC, IPSC, EPSP, IPSP), or 3) learning/memory behavioral task. "No” indicates article looked at and/or measured neuroplasticity but found no change between control and gene (i.e., enhancement or deficit was not indicated in the article). “N/a” indicates no article addressed measured plasticity in this manner.

| Gene/ CNV | Description | PubMed ID | Dendritic morphology | Electro-physiology | Behavior |
|---|---|---|---|---|---|
| *16p11.2* | m7qF3 | 21969575, 31398341, 30089910, 29722793, 29038598, 28984295, 27544825, 26872257, 26572653, 26273832, 25581360, 24794428 | Yes | Yes | Yes |
| *15q13.3* | m7qC | 29395074, 24090792 | Yes | Yes | Yes |
| *15q11-q13* | m7qB5-qC | 19563756, 28691086, 26158416, 25418414, 25144834, 21179543 | Yes | Yes | Yes |
| *17p11.2* | M11qB1.3-qB2 | 22492990, 14709593 | N/A | N/A | Yes |
| *Actl6b* | Actin like 6B | 32312822 | N/A | N/A | Yes |
| *Adnp* | Activity-dependent neuroprotector homeobox | 30106381 | Yes | N/A | Yes |
| *Ahi1* | Abelson helper integration site 1 | 33046712 | N/a | Yes | Yes |
| *Ank2* | Ankyrin 2, neuronal | 31285321 | No | Yes | Yes |
| *Arid1b* | AT-rich interaction domain 1B | 32398858, 29184203, 28867767 | Yes | Yes | Yes |
| *Arx* | Aristaless related homeobox | 29659809 | N/A | Yes | Yes |
| *Auts2* | Activator of transcription and developmental regulator | 32498016 | Yes | Yes | Yes |
| *Cacna1c* | Calcium voltage-gated channel subunit alpha1 C | 16251435, 20573883 | No | Yes | Yes |
| *Camk2a* | Calcium/calmodulin dependent protein kinase II alpha | 28130356 | Yes | Yes | No |
| *Caprin1* | RNA granule protein 105 (Caprin 1) | 26865403 | N/A | N/A | Yes |
| *Cdkl5* | Cyclin-dependent kinase-like 5 | 31201320 | Yes | Yes | Yes |
| *Chd2* | Chromodomain helicase DNA binding protein 2 | 30344048 | N/a | Yes | Yes |
| *Chd8* | Chromodomain helicase DNA binding protein 8 | 27602517, 33837267, 33826902, 30104731, 28671691, 28402856, 27694995 | Yes | Yes | Yes |
| *Cntnap2* | Contactin associated protein-like 2 | 33046712, 30679017, 26273832, 26158416, 21962519 | Yes | Yes | Yes |
| *Ctnnb1* | Catenin (cadherin associated protein), beta 1 | 27131348 | N/A | N/A | Yes |
| *Cul3* | Cullin 3 | 31780330, 31455858 | Yes | Yes | No |
| *Deaf1* | Transcription factor | 24726472 | N/A | N/A | Yes |
| *Dip2a* | Disco interacting protein 2 homolog A | 31600191 | Yes | Yes | No |
| *Dlg4* | Discs, large homolog 4 (Drosophila) | 20952458 | Yes | N/A | Yes |
| *Dmd* | Dystrophin | 19649270 | Yes | Yes | Yes |
| *Dscam* | DS cell adhesion molecule | 23175819 | Yes | N/A | N/A |
| *Dyrk1a* | Dual specificity tyrosine-Y-phosphorylation regulated kinase 1A | 30831192, 11555628, 12192061 | Yes | N/A | Yes |
| *Fmr1* | Fragile X mental retardation protein 1 | 9144249, 16055059,19659572, 21220020,21856714, 22817866, 23083736, 24493647,30679017 | Yes | Yes | Yes |
| *Foxp1* | Forkhead box P1 | 25266127, 28978667 | Yes | Yes | Yes |
| *Foxp2* | Forkhead box P2 | 19490899 | Yes | Yes | N/a |
| *Gabrb3* | Gamma-aminobutyric acid type A receptor beta 3 subunit | 9763493 | N/A | N/A | Yes |
| *Gria2* | Glutamate receptor, ionotropic, AMPA 2 | 12805550 | N/A | Yes | N/A |
| *Grin2b* | Glutamate receptor, ionotropic, N-methyl D-aspartate 2B | 20357110 | Yes | Yes | Yes |
| *Iqsec2* | IQ motif and Sec7 domain 2 | 30842726 | N/A | Yes | Yes |
| *Kmt2a* | Lysine methyltransferase 2A | 27485686 | Yes | Yes | Yes |
| *Mecp2* | Methyl-CpG binding protein 2 | 19812320, 20138994,20163734, 22378884 | Yes | Yes | Yes |
| *Mef2c* | Myocyte Enhancer Factor 2C | 18599438, 27779093 | Yes | Yes | Yes |
| *Mtor* | Mechanistic target of rapamycin kinase | 32711953 | Yes | Yes | N/A |
| *Nbea* | Neurobeachin | 19723784, 23153818, 32994505 | Yes | Yes | Yes |
| *Nf1* | Neurofibromin 1 | 25242307, 11279521, 4213300 | N/A | Yes | Yes |
| *Nlgn3* | Neuroligin 3 | 17823315, 21808020, 22983708, 23761734, 25144834, 26291161, 28921757 | Yes | Yes | Yes |
| *Nr2f1* | Nuclear receptor subfamily 2 group F member 1 | 32320667 | N/A | Yes | Yes |
| *Nr3c2* | Nuclear receptor subfamily 3, group C, member 2 | 16368758 | No | No | Yes |
| *Nrxn1* | Neurexin 1 | 19822762 | N/a | Yes | Yes |
| *Nrxn2* | Neurexin2 | 25745399 | No | Yes | No |
| *Pogz* | Pogo transposable element derived with ZNF domain | 32103003 | Yes | Yes | Yes |
| *Ptchd1* | Patched domain containing 1 | 28416808 | Yes | Yes | Yes |
| *Pten* | Phosphatase and tensin homolog | 23487788, 25288137, 25609613, 31240311 | Yes | Yes | Yes |
| *Rai1* | Retinoic acid induced 1 | 30275311 | Yes | N/A | N/A |
| *Reln* | Reelin | 11580894 | Yes | N/A | N/A |
| *Rims1* | Regulating synaptic membrane exocytosis 1 | 11797009, 15066271, 29891949 | No | Yes | Yes |
| *Rps6ka3* | Ribosomal protein s6 kinase, 90kDa, polypeptide 3 | 23742761 | Yes | Yes | Yes |
| *Scn1a* | Sodium channel, voltage-gated, type 1, alpha subunit | 31870807, 22914087, 26017580, 32126198 | N/A | Yes | Yes |
| *Scn2a* | Sodium voltage-gated channel alpha subunit 2 | 29867081 | N/A | N/A | Yes |
| *Setd5* | SET domain containing 5 | 30455454, 30655503 | Yes | Yes | Yes |
| *Shank2* | SH3 and multiple ankyrin repeat domains 2 | 22699620, 27544825, 27903723 | Yes | Yes | Yes |
| *Shank3* | SH3 and multiple ankyrin repeat domains 3 | 21167025, 21558424, 23621888, 24259569, 26134648, 26559786, 26886798, 27161151, 27492494, 28753255, 30610205, 32199104 | Yes | Yes | Yes |
| *Slc1a2* | Solute carrier family 1 (glial high affinity glutamate transporter), member 2 | 25662838 | N/A | Yes | N/A |
| *Slc6a1* | Solute carrier family 6 member 1 | 20016099 | No | Yes | Yes |
| *Slc9a6* | Solute carrier family 9, subfamily A | 4035762 | Yes | N/A | N/A |
| *Syngap1* | Synaptic Ras GTPase activating protein 1a | 12598599, 22700469, 23141534, 31680851, 32294447 | Yes | Yes | Yes |
| *Tbr1* | T-box, brain 1 | 24441682, 32294447, 31680851, 30792833, 30318412, 25981743, 24441682 | Yes | Yes | Yes |
| *Tbx1* | T-box 1 | 21908517 | N/A | N/A | Yes |
| *Tcf4* | Transcription factor 4 | 27568567, 30705426 | Yes | Yes | Yes |
| *Trio* | Trio Rho guanine nucleotide exchange factor | 30840899 | Yes | Yes | N/A |
| *Tsc1* | Tuberous sclerosis 1 | 25155956, 32375878 | Yes | N/A | Yes |
| *Tsc2* | Tuberous sclerosis 2 | 25155956, 30679017 | Yes | Yes | N/A |
| *Tshz3* | Teashirt zinc finger homeobox 3 | 27668656 | N/A | Yes | N/A |
| *Ube3a* | Ubiquitin protein ligase E3A | 9808466, 19430469, 20211139, 21974935, 22381732, 26166566 | Yes | Yes | Yes |
| *Upf3b* | Regulator of nonsense mediated mRNA decay | 28948974 | Yes | N/A | Yes |